\documentclass[twocolumn,noshowpacs,amsmath,amssymb]{revtex4}

\usepackage{amsmath}
\usepackage[dvipsnames]{xcolor}
\usepackage{graphicx}
\usepackage{dcolumn}
\usepackage{bm}

\newif\ifkeepremark
\keepremarktrue

\newcommand{\gt}{\tilde{g}}
\newcommand{\am}{a}
\newcommand{\ppm}{\mathbf{p}}
\newcommand{\M}{\mathbf{M}}
\newcommand{\Mb}{\overline{\mathbf{M}}}
\newcommand{\F}{\mathbf{F}}
\newcommand{\Q}{\mathbf{Q}}

\newcommand{\I}{\mathbf{I}}

\newcommand{\ONEm}{\mathbf{1}}

\newcommand{\Am}{\mathbf{A}}
\newcommand{\Pm}{\mathbf{P}}
\newcommand{\Dm}{\mathbf{D}}

\begin{document}


\title{Mortality equation characterizes the dynamics of an aging population}

\author{T. Fink}
\affiliation{
London Institute for Mathematical Sciences, Royal Institution, 21 Albermarle St, London W1S 4BS, UK
}

\begin{abstract}
\noindent
Aging is thought to be a consequence of intrinsic breakdowns in how genetic information is processed.
But mounting experimental evidence suggests that aging can be slowed.
To help resolve this mystery, I derive a mortality equation which characterizes the dynamics of an evolving population with a given maximum age.
Remarkably, while the spectrum of eigenvalues that govern the evolution depends on the fitness, how they change with the maximum age is independent of fitness.
This makes it possible to establish the conditions under which programmed aging can provide an evolutionary benefit.
\end{abstract}

\keywords{}
\maketitle
\begin{figure}[b!]
\includegraphics[width=1\columnwidth]{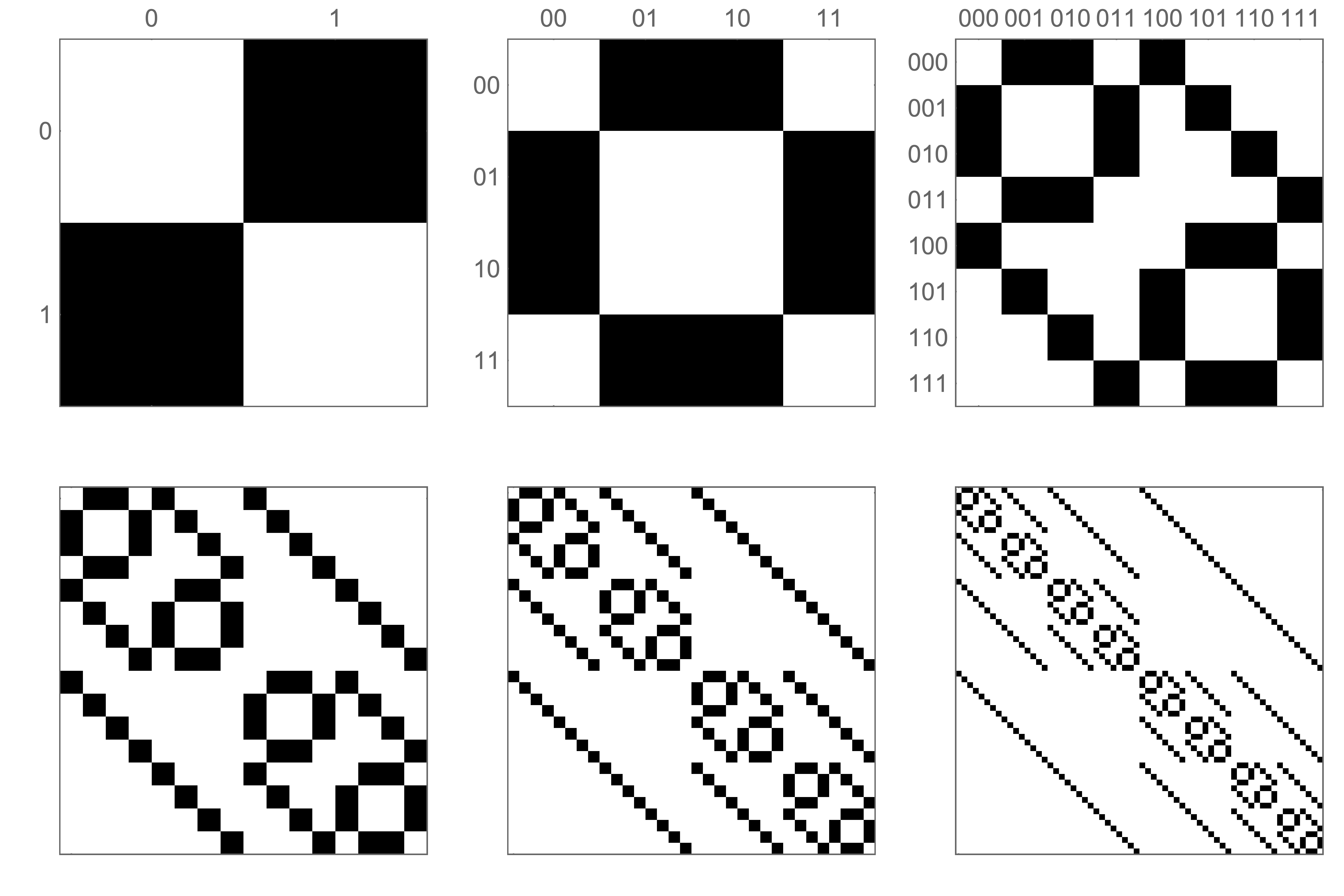}
\caption{
    \textbf{The first nine unnormalized mutation matrices $\Mb$.} 
    These $2^n \times 2^n$ matrices act to diffuse a population across genotype space via point mutations.
    They correspond to genomes of size $n$, from $n=1$ to $9$, where white and black indicate 0 and 1.
    The normalized mutation matrix is $\M = \Mb/n$, so that the columns and rows all add up to one.
}
\vspace{-0.1in}
\label{MMatrices}
\end{figure}
\begin{figure}[b!]
    	\centering
    	\includegraphics[width=1\columnwidth]{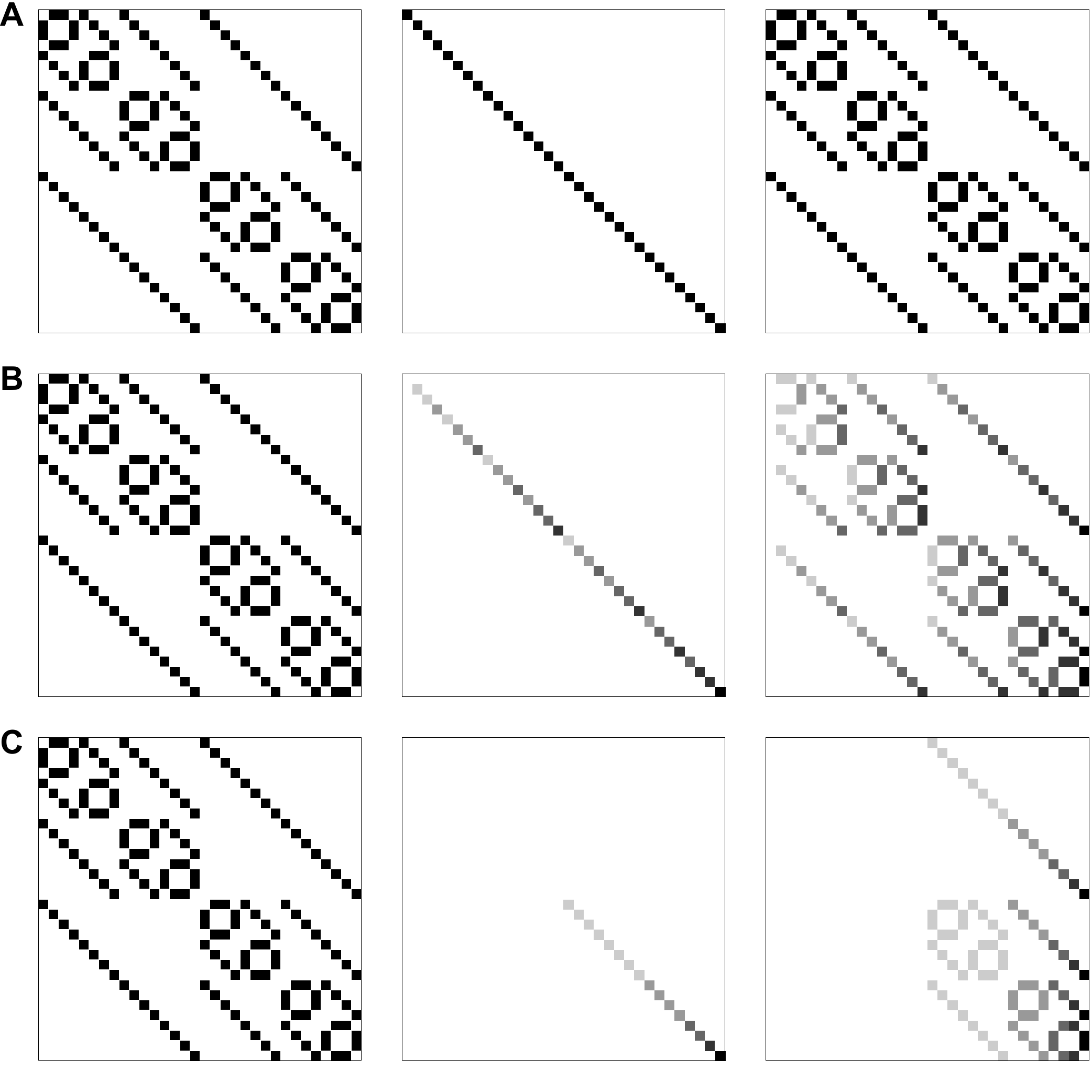}
    	\caption{
	\textbf{The mutation matrix $\M$, the fitness matrix $\F$ and their product $\M \F$ for different fitness functions.}
	In all cases the genome length is $n=5$.
	\textbf{A} 	The fitness is constant, equal to 1 for all genotypes.
	\textbf{B} 	The fitness is the fraction of of bits at which the genotype and the target genotype $11111$ match.
	\textbf{C} 	The fitness is the overlap fitness: 
			the length of the longest initial substring over which the genotype and the target genotype $11111$ match, divided by $n$.
    }
    \label{MFMF}
\end{figure}
\begin{figure}[t!]
    	\centering
    	\includegraphics[width=1\columnwidth]{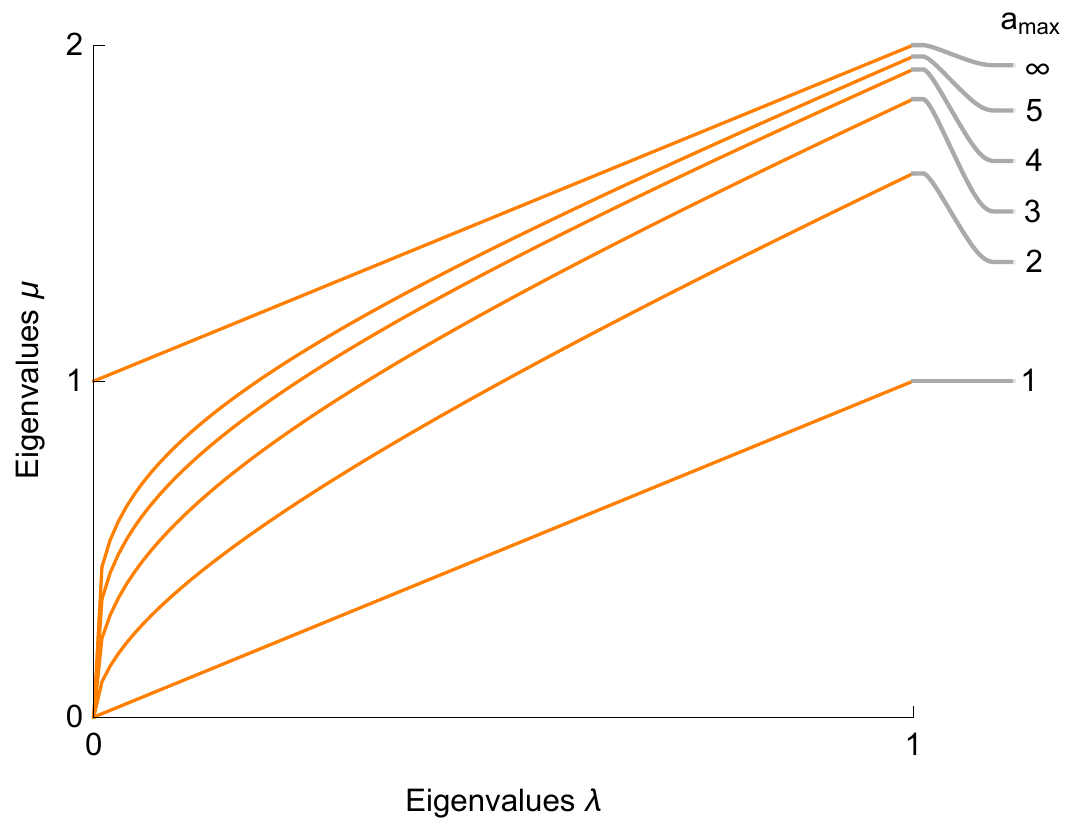}
    	\caption{
	\textbf{Eigenvalues for different maximum ages.}
	For any eigenvalue $\lambda$ of the transition matrix $\M\F$ ($\am=1$), the eigenvalue $\mu$ of the transition matrix $\Q$ for higher $\am$ is given by jumping up to the appropriate line.
    	The fitness $\F$ alone determines the horizontal placement of the eigenvalues, but $\am$ alone determines their vertical placement.
    }
    \label{EigenvalueTransform}
\end{figure}
\begin{figure*}[t!]
    \centering
    \includegraphics[width=1 \textwidth]{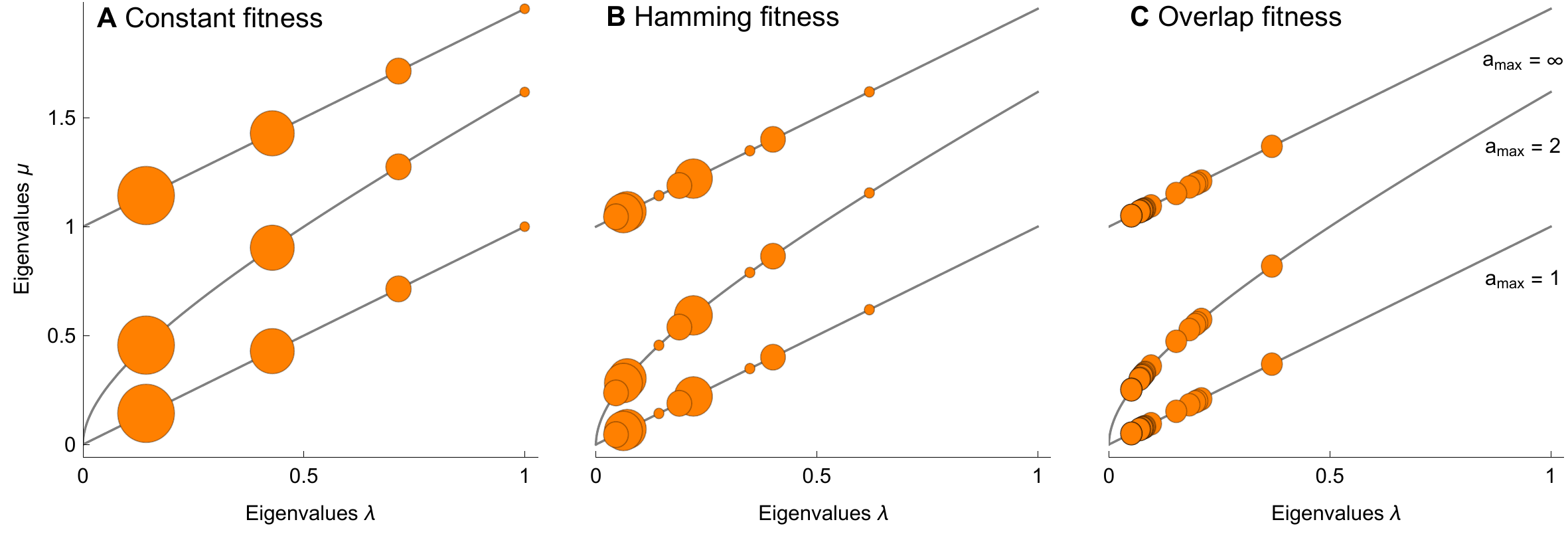}
    \vspace{-0.15in}
    \caption{
    	\textbf{Testing our theory with three fitness functions.}
   	 I tested the prediction of eq. (\ref{Main2}), shown in Fig. \ref{EigenvalueTransform}, 
	 with populations evolving according to three fitness functions: constant, Hamming and overlap fitness.
   	 For each one, I computed the spectrum of positive real eigenvalues for maximum age $\am = 1$, 2 and $\infty$; 
	 all other eigenvalues have modulus $< 1$.
	 The circles are the different eigenvalues and their area is proportional to their degeneracies.
	 The lines are the predictions from eq. (\ref{Main2}).
    }
    \label{FitnessComparison}
\end{figure*}

\noindent 
Why do we get old?
Darwin's theory of evolution is the result of mutation, inheritance and selection.
It doesn't refer to death explicitly, but dying is a consequence of competition for finite resources,
or when one life becomes the resource for another.
Some mutants, ill equipped to cope with their environment, simply break down.
\\ \indent 
This accounts for the pervasiveness of death, but it says nothing about the universality of aging.
The explanation for aging is fraught with debate.
On the one hand, the canonical view is that aging is the result of intrinsic breakdowns in how genetic information is processed \cite{Rozing17,Kirkwood11,Kirkwood16}.
The selective pressure against these breakdowns declines throughout adult lifespan, in line with the future expected reproductive output \cite{Kirkwood16}.
\\ \indent 
On the other hand, many scientists believe that aging is programmed \cite{Travis04,Mitteldorf,Werfel15,Werfel17,Goldsmith08,Goldsmith12,Goldsmith13}
Rather than being a fundamental and unavoidable attribute of biological life, 
aging is actively sought by evolution because it is advantageous.
This group rejects the genetic breakdown argument because, while plausible, we don't know how little or how much breakdown is imposed by the laws of physics.
Biological life is, after all, one instance of a self-replicating machine, 
and our understanding of the thermodynamic constraints on such systems is poor.
Once heretical, programmed aging is gaining traction because of a range of interventions that reverse aging markers \cite{Sinha2014}.
Research into the field is speeding up, thanks to generous funding by longevity companies such as Altos Labs \cite{Regalado21}.
\\ \indent 
Several models have been put forward to support programmed aging \cite{Werfel15}, which were reviewed in detail by Kowald and Kirkwood \cite{Kirkwood16}.
These models ``have relied extensively on simulation techniques rather than on mathematical analysis. 
While analytical (mathematical) models generally have the advantage of clarity, 
they quickly become intractable when the phenomenon to be analysed depends on features such as spatial effects, 
which are at the heart of several of the claims made in favor of programmed aging'' \cite{Kirkwood16}.
Spatial effects, while part of any physical manifestation of life, can have unforeseen consequences, 
such as kin selection, whereby reproductive benefit is transferred to relatives.
This can cloud more basic questions about the evolutionary benefit of aging.
\\ \indent 
In this Letter I provide a simple mathematical framework for answering fundamental questions about the evolutionary benefits of programmed aging. 
My model is independent of the choice of fitness function and does not invoke spatial effects.
I purposefully do not address how a population might evolve from an immortal one to a mortal one, which concerns the challenge of group selection.
Potential mechanisms for doing so include violating mean-field assumptions commonly used in evolutionary biology \cite{Werfel15}.
Rather, I want to settle a more fundamental question: from an evolutionary perspective, is aging a trait worth fighting for in the first place? 
If programmed aging is favored by natural selection, rather than being an inevitable consequence of genetic breakdowns, 
then it may be possible to devise genetic and pharmacological interventions to prolong life and combat disease.
\\ \indent 
In this Letter I do four things.
First, I derive a simple mortality equation that governs the transition matrix $\Q$ of an evolving population with maximum age $\am$.
It is
\begin{equation*}
\Q^{\am}(\I + \M \F - \Q) = \M \F,
\end{equation*}
where 
$\M$ is the mutation matrix shown in Fig. \ref{MMatrices} and 
$\F$ is a diagonal matrix with the genotype fitnesses along the diagonal. 
This equation is remarkable because it applies for any fitness function.
Second, I show how to solve the mortality equation in terms of the spectrum of eigenvalues of the transition matrix, which completely determines the dynamics. 
I provide a geometric characterization of how these eigenvalues change as the maximum age $\am$ increases from 1 to infinity (immortality), shown in Fig. \ref{EigenvalueTransform}.
The solution has a very special property: while the spectrum of eigenvalues depends on the fitness, how they change with $\am$ is independent of the fitness.
Third, I test the solution to the mortality equation by numerically solving it for three specific fitness functions: constant fitness, Hamming fitness and overlap fitness.
The spectrum of eigenvalues for these evolving populations perfectly agrees with those given by the solution, as shown in Fig. \ref{FitnessComparison}.
Fourth, I show that, in a constant environment, programmed aging is not favored by natural selection.
But in a changing environment, in which the equilibration time becomes important, programmed aging can provide an evolutionary benefit.
\\ \\ \noindent {\sf\textbf{\textcolor{purple}{Mutation}}}
\\ In my model, we have a population of reproducing individuals with binary genotypes of length $n$.
Thus there are $2^n$ different possible genotypes. 
For example for $n=3$, these are $000, 001, 010, \ldots, 111$.
The population vector is given by $\mathbf{p} = (p_1, \ldots, p_{2^n})$,
which is the size of the subpopulations with genotypes $1, \ldots, 2^n$.
For $n=3$, 
$p_1$ is size of the subpopulation with genotype $000$, 
$p_2$ is size of the subpopulation with genotype $001$, 
and so on.
Offspring are identical to their parents apart from a single point mutation in the genotype, that is, each child has one spelling mistake.
\\ \indent
The process of mutation in the population can be captured by the mutation matrix $\Mb$, shown in Fig. \ref{MMatrices}.
The rows of $\Mb$ correspond to genotype $00\ldots$ at the top to genotype $11\ldots$ at the bottom (see Fig. \ref{MMatrices} top).
The same applies to the columns from left to right.
The ones in each row indicate the different parents that can give birth to that row's genotype,
keeping in mind that an offspring differs from its parent by a single bit.
For example, for $n=3$, the child genotype 001 can arise from mutations in the parents 000, 011 and 101, 
since these are the only genotypes that are one bit away.
\\ \indent 
Eventually we will want our mutation matrix to be normalized, meaning that all the rows and columns add to one.
But for now it is convenient to work with the unnormalized mutation matrix $\Mb$, 
and later normalize it by just dividing $\Mb$ by $n$ to get $\M$.
The mutation matrix $\Mb$ can be defined recursively in block form:
\begin{equation}
\Mb_{n+1} =
\left(
\begin{array}{cc}
  \Mb_n & \I_n    \\ 
   \I_n & \Mb_n     
\end{array}
\right)\!, 
\quad \!
\Mb_1 = \left(
\begin{array}{cc}
	0 & 1  \\
 	1 & 0    
\end{array}\right)\!,
\label{BN}
\end{equation}
where $\Mb_n$ and $\I_n$ are the $2^n\times 2^n$ mutation matrix and identity matrix.
We can find the eigenvalues of $\Mb$ as follows.
Let $p_{\Mb_n}$ be the characteristic polynomial of $\Mb_n$, the roots of which are the eigenvalues of the matrix.
Then
\begin{eqnarray*}
p_{\Mb_{n+1}}(\lambda) &=& \det \left(\begin{array}{cc}
	\lambda \I_n - \Mb_n & - \I_n    \\ 
	-\I_n              & \lambda \I_n - \Mb_n     
\end{array} \right) \\
    &=& \det(\Mb_n - (\lambda-1) \I_n) \cdot \det(\Mb_n - (\lambda+1) \I_n)     \\
    &=& p_{\Mb_n}(\lambda)|_{\lambda=\lambda-1} \cdot p_{\Mb_n}(\lambda)|_{\lambda=\lambda+1},
\end{eqnarray*}
\noindent where $p_{\Mb_0} = \lambda$.
Thus the characteristic polynomial $p_{\Mb_{n+1}}$ is the product of $p_{\Mb_n}$ evaluated at $\lambda-1$ and $p_{\Mb_n}$ evaluated at $\lambda+1$.
This recursive step can be understood visually though a Pascal's triangle of terms:
the product of the terms in row $n$ is the characteristic polynomial $p_{\Mb_n}$, where the rows start at 0:
\begin{equation}
\arraycolsep = -1.3pt
\begin{array}{ccccccccc}
        	 		&               	&               		&               		& \lambda		&               		&               		&              		&          		\\
  	          	&               	&               		& \lambda - 1		&               	& \lambda + 1 		&               		&               	&        		\\ 
            		&               	& \lambda - 2     	&               		& \lambda^2 	&              	 		& \lambda + 2  		&               	&        		\\
            		& \lambda - 3	&               		& (\lambda - 1)^3 	&               	& (\lambda + 1)^3 	&               		& \lambda + 3	&   			\\            
\lambda - 4	& 			& (\lambda - 2)^4 	&               		& \lambda^6 	&               		& (\lambda + 2)^4	&               	& \lambda + 4.
\end{array}
\label{JA}
\end{equation}
\noindent We immediately see that the eigenvalues of $\Mb_n$ are 
   $n, n-2, n-4, \ldots, -n$, 
with degeneracies 
$\binom{n}{0}, \binom{n}{1}, \ldots, \binom{n}{n}$.
So the eigenvalues of $\M_n = \Mb_n/n$ are 
$1, 1-\frac{2}{n}, 1-\frac{4}{n}, \ldots, -1$, 
with the same degeneracies.
The principal eigenvector of $\M_n$ is $(1,1,1,\ldots)$.
 \\ \\ {\sf\textbf{\textcolor{purple}{Selection}}}
\\ 
An environment amounts to an assignment of a fitness to each genotype, where genotypes with the same fitness are said to have the same phenotype.
To give preferential treatment to certain genotypes, we need to define a fitness function.
The fitness of each individual is determined by the distance of the genotype $g$ from some target genotype $\tilde{g}$ which is optimally suited to the environment.
The closer $g$ is to the optimum $\tilde{g}$, the higher is the individual's fitness, and therefore reproduction rate, which is proportional to fitness.
As the population reproduces and mutates, it drifts towards the vicinity of this optimum $\tilde{g}$.
The way in which it does so critically depends on the kind of distance that is used.
In principle any fitness can be related to a distance function, even if it is just a lookup table.
\\ \indent 
Let $\mathbf{f} = (f_1, \ldots, f_{2^n})$ be the vector of fitnesses for the $2^n$ genotypes, where $f_i \in [0,1]$.
Then the fitness matrix $\F$ is the square matrix with $\mathbf{f}$ along the diagonal but otherwise zero. 
\\ \indent
Let the maximum age $\am$ be the number of times that an individual reproduces before dying.
Let's assume, for now, that $\am = 1$.
Then the distribution of the population at time $t$ gets transformed to an updated distribution at time $t+1$ according to the transition matrix $\M\F$:
\begin{equation}
\ppm(t+1) = \M\F \ppm(t).
\end{equation}
While the key results in this Letter are independent of the specific fitness function $\F$, it is illustrative to consider actual examples.
In Fig. \ref{MFMF} I show the matrices $\M$, $\F$ and $\M\F$ for three different fitness functions: constant, Hamming and overlap fitness.
Later we will use these to test our theory.
\\ \indent 
Starting from a given initial population at time $t=0$, we can determine the population distribution at time $t$ by repeatedly applying the matrix $\M\F$, or just raising it to a power:
\begin{equation*}
\ppm(t) = (\M\F)^t \ppm(0).
\end{equation*}
The steady state distribution, which corresponds to $t = \infty$, 
is given by the principal eigenvector of the matrix $\M\F$, and the long term growth rate is given by the largest real eigenvalue of $\M\F$.
\\ \\ \noindent {\sf\textbf{\textcolor{purple}{The mortality equation}}}
\\ The matrix $\M\F$ tells us how a population with maximum age $\am=1$ evolves. 
What we really want, however, is to understand how a population evolves for an arbitrary $\am$.
I now derive a matrix polynomial which governs the transition matrix $\Q$ for a population with $\am \geq 1$.
\\ \indent  
Let's start by considering $\am=2$.
Let $\mathbf{x_1}$ be the size of the subpopulation with age 1 and genotypes $1, \ldots, 2^n$, and
$\mathbf{x_2}$ be the size of the subpopulation with age 2 and genotypes $1, \ldots, 2^n$.
Individuals with ages 1 and 2 can give birth, but all offspring are born with age 1.
Let $\mathbf{p} = \mathbf{x_1} + \mathbf{x_2}$ be the total population size.
Then
\begin{equation}
    \mathbf{x_1}(t+1) = \M\F \mathbf{p}(t) \quad {\rm and} \quad \mathbf{x_2}(t+1) = \mathbf{x_1}(t).
    \label{M1}
\end{equation}
Inserting these into $\mathbf{p}(t+1) = \mathbf{x_1}(t+1) + \mathbf{x_2}(t+1)$ gives
\begin{equation*}
    \mathbf{p}(t+1) =  \M\F \mathbf{p}(t) + \mathbf{x_1}(t).
\end{equation*}
Incrementing both sides by one time step and again applying eq. (\ref{M1}) gives
\begin{equation}
    \mathbf{p}(t+2) =  \M\F \mathbf{p}(t+1) + \M\F \mathbf{p}(t).
    \label{M5}
\end{equation}
Our goal is to obtain the transition matrix $\Q$ for which $\mathbf{p}(t+1) = \Q \mathbf{p}(t)$.
Writing (\ref{M5}) in terms of $\Q \mathbf{p}(t)$, we find
$\Q^2 \mathbf{p}(t) = \M\F (\Q + \I)\mathbf{p}(t)$, 
and so $\Q$ satisfies 
\begin{equation*}
\Q^2 = \M\F (\I + \Q). 
\end{equation*}
\indent 
We can take a similar approach for general maximum age $\am$.
Now we need to keep track of $\am$ subpopulation vectors, with ages $1, \ldots, \am$.
We use $\mathbf{x_1}, \ldots, \mathbf{x_{\am}}$ to indicate their vectors.
Individuals of all ages can give birth, but all offspring are born with age 1.
Let 
\begin{equation}
\mathbf{p}(t) = \mathbf{x_1}(t) + \ldots + \mathbf{x_{\am}}(t)
\label{LB}
\end{equation}
be the total population size, of all ages.
Then
\begin{eqnarray}
    \mathbf{x_1}(t+1) = \M\F \mathbf{p}(t) \quad {\rm and} \quad
    \mathbf{x_{i+1}}(t+1) = \mathbf{x_i}(t).
    \label{LD}
\end{eqnarray}
Inserting these into eq. (\ref{LB}) evaluated at time $t+1$ gives
\begin{equation*}
    \mathbf{p}(t+1) =  \M\F \mathbf{p}(t) + \mathbf{x_1}(t) + \ldots \mathbf{x_{\am-1}}(t).
\end{equation*}
Incrementing the time by 1 and again applying eq. (\ref{LD}),
\begin{equation*}
    \mathbf{p}(t+2) =  \M\F \mathbf{p}(t+1) + \M\F \mathbf{p}(t) + \mathbf{x_1}(t) + \ldots \mathbf{x_{\am-2}}(t).
\end{equation*}
Repeating this process until all of the $\mathbf{x}$s are converted to $\mathbf{p}$s,
we obtain
\begin{equation*}
    \mathbf{p}(t+\am) =  \M\F (\mathbf{p}(t + \am - 1) + \ldots + \M\F (\mathbf{p}(t).
\end{equation*}
Then, with $\mathbf{p}(t+1) = \Q \mathbf{p}(t)$, $\Q$ satisfies 
\begin{equation}
    \Q^{\am} = \M\F (\I + \Q + \ldots + \Q^{\am-1}). 
    \label{LO}
\end{equation}
Since, for a general matrix $\Am$,
\begin{equation*}
\I + \Am + \ldots + \Am^{\am-1} = (\I - \Am)^{-1} (\I-\Am^{\am}),
\end{equation*}
we can write
	$(\I - \Q) \Q^{\am} = \M\F \left(\I - \Q^{\am}\right)$
or, equally,
\begin{equation}
	\Q^{\am} (\I + \M\F - \Q) = \M\F.
	\label{Main1}
\end{equation}
I call eq. (\ref{Main1}) the mortality equation, and it is one of the two main results of this Letter.
Its concision belies its power.
It gives the transition matrix of a population with maximum age $\am$ in terms of the transition matrix with maximum age 1, for any fitness function $\F$.
Notice that while the compact eq. (\ref{Main1}) has degree $\am+1$, 
it can always be reduced to degree $\am$ by dividing through by $\I - \Q$ to give eq. (\ref{LO}).
\\ \\ \noindent {\sf\textbf{\textcolor{purple}{Solving the mortality equation}}} \\
With our mortality equation in hand, what kind of meaning can we extract from it?
It is unusual in that it is a polynomial equation in terms of matrices rather than numbers:
$\I$ can be thought of as one and $\M\F$ as a constant, and we are solving the polynomial for $\Q$.
However, matrix polynomials are subject to most of the same machinery used to solve ordinary polynomials, with some extra care required for taking roots.
\\ \indent 
Taking roots of a matrix works as follows.
A matrix $\Am$ is diagonalizable if there exists an invertible matrix $\Pm$ and a diagonal matrix $\Dm$ such that $\Am = \Pm \Dm \Pm^{-1}$.
The roots of such an $\Am$ are then $\sqrt[i]{\Am} = \Pm \sqrt[i]{\Dm} \Pm^{-1}$,
where $\sqrt[i]{\Dm}$ just amounts to taking the roots of the diagonal elements, bearing in mind to take all $i$ roots of each.
\\ \indent 
To gain some intuition for our mortality equation, 
let's consider the simple case of constant fitness, described below, in which all genotypes have fitness 1.
Then $\F = \I$ and $\M \F = \M$.
Applying $\M$ to a population has the effect of diffusing it over the hypercube, smoothing it out towards a uniformly distributed population. 
Eq. (\ref{Main1}) then becomes $\Q^{\am} (\I + \M - \Q) = \M$.
For maximum age $\am=2$, this can be solved like an ordinary quadratic:
\begin{equation}
	\Q = \M/2 \pm\sqrt{(\M/2)^2+\M}.
	\label{MD}
\end{equation}
As we shall see, only one such matrix has positive real eigenvalues.
The specific matrices depend on $n$. 
For $n=2$,
\begin{eqnarray}
	\Q = \textstyle \M/2 + \left(\left(\sqrt{5} + \sqrt{3}i\right) \! \ONEm - 4 \sqrt{3} i \M\right)\!/8,
	\label{NA}
\end{eqnarray}
where $\ONEm$ is the all-1s matrix.
The matrix $\Q$ has eigenvalues $\frac{1+\sqrt{5}}{2}, 0, 0, \frac{-1+\sqrt{3} i}{2}$. 
Notice that, even though $\M$ and $\F$ are real, $\Q$ and its eigenvalues can be complex.
\\ \indent 
We can in principle extend this process to solve for larger genome size $n$ and maximum age $\am$.
But this requires painstaking matrix inversion and is not analytically tractable.
Fortunately, there is a shortcut for computing the eigenvalues of $\Q$ from those of $\M\F$.
By the Cayley Hamilton theorem, the eigenvalues of $\Q$ have the same functional relation to the eigenvalues of $\M \F$ as the matrix $\Q$ does to the matrix $\M \F$.
Let $\mu_1, \ldots, \mu_{2^n}$ be the eigenvalues of $\Q$.
We can express them in terms of $\lambda_1, \ldots, \lambda_{2^n}$, the eigenvalues of $\M\F$,
using the analogue to eq. (\ref{Main1}), but for numbers rather than matrices:
\begin{equation}
	\mu^{\am} (1 + \lambda - \mu) = \lambda.
	\label{Main2}
\end{equation}
This is the second of the two main results of this Letter.
Notice that, as for the matrix $\Q$ in eq. (\ref{Main1}), this has degree $\am+1$.
But it can always be reduced to degree $\am$ by dividing through by $\mu - 1$ to give $\mu^{\am} = \lambda (1 + \mu + \ldots + \mu^{\am-1})$. 
Then the solution $\mu=1$ to eq. (\ref{Main2}) is spurious, and we can neglect it in what follows.
\\ \indent
Writing eq. (\ref{Main2}) as $\mu^{\am} = \lambda/(1 + \lambda - \mu)$, and recalling $\lambda \leq 1$, 
we see that the equation has only one positive real root, disregarding $\lambda = 1$. 
This is because the monomial on the left and the shifted hyperbola on the right intersect at 1 and at a single point between 0 and 1.
\\ \indent
The positive real solutions of eq. (\ref{Main2}) are
For $\am=1$, eq. (\ref{Main2}) reduces to $\mu = \lambda$.
This is expected, since for $\am=1$ the transition matrix $\Q$ is $\M\F$.
This eigenvalue relation is shown as the bottom line in Fig. \ref{EigenvalueTransform}.
For $\am=\infty$, we can also solve for $\mu$ explicitly.
We know that $\mu > 1$, because otherwise $\mu^{\am}$ in eq. (\ref{Main2}) vanishes but $1 + \lambda - \mu$ is finite.
So $\mu^{\am}$ is infinite, meaning $1 + \lambda - \mu$ must vanish, that is, $\mu = \lambda + 1$.
This is shown as the top line in Fig. \ref{EigenvalueTransform}.
For intermediate values of $\am$, the positive eigenvalues $\mu$ are between these two lines.
For $\am=2$,
\begin{equation}
	\mu = \lambda/2 + \sqrt{(\lambda/2)^2+ \lambda}. 
	\label{NK}
\end{equation}
This is the second line from the bottom in Fig. \ref{EigenvalueTransform}.
For $\am=3$, the eigenvalues $\mu$ are given by the third line from the bottom.
As $\am$ grows, the lines are successively higher, ultimately converging to $\mu = \lambda + 1$.
\\ \\ \noindent {\sf\textbf{\textcolor{purple}{Testing our theory with three fitness functions}}}
\\ To test eq. (\ref{Main2}) and the solution shown in Fig. \ref{EigenvalueTransform}, 
I wrote a program to explicitly compute the transition matrices $\Q$ for three very different fitness functions: 
constant fitness, Hamming fitness and overlap fitness, described below.
I then computed the spectrum of eigenvalues of these matrices and found that they agree exactly with the prediction.
This confirmation is shown in Fig. \ref{FitnessComparison},
where for each fitness function, I show the spectrum for maximum age $\am = 1$, 2 and $\infty$.
\\ \indent
First, I considered constant fitness, in which every genotype has fitness 1 ($\F = \I$).
Constant fitness diffuses the population over the hypercube, smoothing it out towards a uniformly distributed population.
The spectrum of eigenvalues for constant fitness is shown in Fig.  \ref{FitnessComparison}A.
As a reality check, we know from eq. (\ref{JA}) that the eigenvalues of $\M\F = \M$ for $n=2$ are $1,0,0,-1$.
Plugging these into eq. (\ref{NK}) gives the same eigenvalues as we obtained by brute force below eq. (\ref{NA}).
\\ \indent 
Second, I considered the Hamming fitness $h$. 
This corresponds to a natural notion of distance: 
the number of edges on the hypercube that must be traversed to get from one corner to another.
The Hamming fitness is one minus $1/n$ times the Hamming distance between some genotype $g$ and the optimal genotype $\tilde{g}$, that is, 
the fraction of bits where $g$ and $\tilde{g}$ match.
For example, if $\tilde{g}$ is 01100 and $g$ is $11000$, then $h = 3/5$.
The spectrum of eigenvalues is shown in Fig.  \ref{FitnessComparison}B.
\\ \indent 
Third, I considered the overlap fitness $v$. 
This is $1/n$ times the length of the longest initial substring over which some genotype $g$ and the optimal genotype $\gt$ match.
For example, if $\tilde{g}$ is 01100 and $g$ is $01010$, then the longest initial overlap is 2, 
since the first two bits of $g$ and $\gt$ match but not the first three, and $v = 2/5$.
The overlap fitness is unusual in that it allows jump discontinuities:
one mutation in the first part of the string can change the fitness from high to low or low to high.
The spectrum of eigenvalues is shown in Fig.  \ref{FitnessComparison}C.
\\ \indent 
The most striking aspect of the mortality equation is its universality.
It tells us how the spectrum of eigenvalues changes, independent of the choice of fitness function.
This is important, because in general quantitative claims about evolving systems assume a specific fitness function, 
whereas the claims here apply to all possible ones.
The fitness matrix $\F$ alone governs the horizontal placement of the eigenvalues in Fig. \ref{FitnessComparison},
whereas the maximum age $\am$ alone governs their vertical placement.
Shifting $\am$ causes the eigenvalues to jump up and down vertically between levels.
\\ \\ \noindent {\sf\textbf{\textcolor{purple}{Aging}}}
\\ The mortality equation provides a framework for gaining quantitative insights into aging, without having to specify the fitness function or invoke spatial effects.
In particular, it can tell us under what conditions programmed aging provides an evolutionary advantage.
\\ \indent 
In a fixed environment, the long-term growth rate of our population is set by the largest eigenvalue $\mu$ of $\Q$.
As Figs. \ref{EigenvalueTransform} and \ref{FitnessComparison} illustrate, this increases with the maximum age $\am$, from $\mu = \lambda$ at $\am=1$ to $\mu = \lambda+1$ at $\am = \infty$.
In the long run, no matter what the fitness function is, a population with $\am = 1$ grows more slowly than one with $\am = 2$,
which in turn grows more slowly than one with $\am = 3$, and so on.
The fastest growing population is the immortal one.
In a constant environment, there is no growth rate benefit afforded by aging; mortality is a losing strategy and programmed aging is not favored by natural selection.
\\ \indent 
However, in reality populations do not reach steady state, where they are optimally suited to their environment.
Rather, they are continually out of equilibrium, slowly adapting to a slowly changing environment.
In this case the equilibration time plays an important role, because the ability to adapt quickly can overcompensate for a slower growth rate.
This is analogous to how, in a race around Britain, a slower yacht can still win by tacking faster.
It is well known that the rate of convergence to equilibrium is controlled by the spectral gap \cite{Pinsky05}.
For a discretely evolving system like ours, the spectral gap is $\mu_1/\mu_2$, where $\mu_1$ and $\mu_2$ are the first and second eigenvalues.
If the spectral gap increases as $\am$ decreases, then in a changing environment programmed aging can provide an evolutionary advantage.
\\ \indent 
Let's compare the spectral gap $\mu_1/\mu_2$ for  $\am = 1$, $\am = 2$ and  $\am = \infty$.
The gaps are, respectively, 
\begin{equation*}
\textstyle
\frac{\lambda_1}{\lambda_2}, \quad
\frac{\lambda_1}{\lambda_2}  \frac{1+\sqrt{1 + 4/\lambda_1}}{1+\sqrt{1 + 4/\lambda_2}} \quad {\rm and} \quad
\frac{\lambda_1+1}{\lambda_2 + 1}.
\end{equation*}
It is readily confirmed that, as $\am$ decreases, the spectral gap increases.
\\ \indent 
This is a remarkable observation. 
It opens up the possibility that programmed aging, rather than being a thermodynamic necessity, could be evolutionarily beneficial, and therefore susceptible to intervention. 
Extensions of the work presented here, such as a complete solution for a specific fitness function---a kind of ``hydrogen atom'' of the mortality equation---would 
 clarify the specific conditions under which programmed aging is favored by natural selection.
 I hope other theorists will find this a fruitful framework for forging new insights into the phenomenon of aging. 
\\ \indent 
This research was supported by a grant from bit.bio. 
I acknowledge Robert Farr for helpful advice and for introducing the concept of the overlap fitness.



\begin{thebibliography}{1}
\begin{footnotesize}
\bibitem{Rozing17}      	M. Rozing, T. Kirkwood, R. Westendorp,	Is there evidence for a limit to human lifespan?,                                           				Nature           			{\bf 546},	E11 			(2017). 	
\bibitem{Kirkwood11} 	T. Kirkwood, S. Melov,       			On the programmed/non-programmed nature of ageing within the life history,                 		Curr Biol     			{\bf 21}, 	R701 		(2011).	
\bibitem{Kirkwood16}    	A. Kowald, T. Kirkwood,  				Can aging be programmed? A critical literature review								Aging Cell       			{\bf 15}, 	986 			(2016).	
\bibitem{Travis04}        	J. Travis,                              			The evolution of programmed death in a spatially structured population,                     			J Gerontol A-Biol J 		{\bf 59},	B301			(2004). 	
\bibitem{Mitteldorf}    	J. Mitteldorf, A. Martins,	            		Programmed life span in the context of evolvability,                                        				Am Nat,				{\bf 184}, 	289 			(2014). 	
\bibitem{Werfel15}      	J. Werfel, D. Ingber, Y. Bar-Yam,		Programed death is favored by natural selection in spatial systems,                         			Phys Rev Lett	  		{\bf 114}, 	238103 		(2015).	
\bibitem{Werfel17}      	J. Werfel, D. Ingber, Y. Bar-Yam,,  		Theory and associated phenomenology for intrinsic mortality arising from natural selection, 	PLOS ONE,        		{\bf 12}, 	e0173677 	(2017).	
\bibitem{Goldsmith08}   	T. Goldsmith		 				Aging, evolvability, and the individual benefit requirement, 							J Theor Biol, 			{\bf 252}, 	764			(2008). 	
\bibitem{Goldsmith12}   	T. Goldsmith                           			On the programmed/non-programmed aging controversy,                                        			Biochemistry (Mosc.)    	{\bf 77}, 	729 			(2012). 	
\bibitem{Goldsmith13} 	T. Goldsmith                            			Arguments against non-programmed aging theories,                                            			Biochemistry (Mosc.)		{\bf 78}, 	971 			(2013). 	
\bibitem{Sinha2014} 		M. Sinha et al., 						Restoring systemic GDF11 levels reverses age-related dysfunction in mouse skeletal muscle,	Science,				{\bf 344}, 	649 			(2014). 	
\bibitem{Regalado21}    	A. Regalado,  						Meet Altos Labs, Silicon Valley's latest wild bet on living forever,						MIT Tech Rev,     		Sep 4,				2021. 	
\bibitem{Pinsky05}		R. Pinsky,						Spectral gap and rate of convergence to equilibrium for a class of conditioned Brownian motions,	Stoch Proc Appl,		{\bf 115},	875			(2005). 	
\end{footnotesize}
\end{thebibliography}
\end{document}